\documentclass[fleqn,10pt]{wlscirep}
\usepackage[utf8]{inputenc}
\usepackage[T1]{fontenc}
\title{Tunable ferroelectricity in hBN intercalated twisted double-layer graphene}

\author[1,*,+]{Yibo Wang}
\author[1,+]{Siqi Jiang}
\author[1]{Jingkuan Xiao}
\author[1]{Xiaofan Cai}
\author[1]{Di Zhang}
\author[1]{Ping Wang}
\author[1]{Guodong Ma}
\author[1]{Yaqing Han}
\author[1]{Jiabei Huang}
\author[2]{Kenji Watanabe}
\author[2]{Takashi Taniguchi}
\author[1,*]{Alexander S. Mayorov}
\author[1,3,*]{Geliang Yu}
\affil[1]{National Laboratory of Solid State Microstructures and School of Physics, Nanjing University, Nanjing 210093, People’s Republic of China.}
\affil[2]{National Institute for Materials Science, 1-1 Namiki, Tsukuba 305-0044 Japan.}
\affil[3]{Collaborative Innovation Centre of Advanced Microsctructures, Nanjing University, Nanjing 210093, People’s Republic of China.}

\affil[*]{yibowang@nju.edu.cn, mayorov@nju.edu.cn, yugeliang@nju.edu.cn}

\affil[+]{These authors contributed equally to this work.}


\begin{abstract}
Van der Waals (vdW) assembly of two-dimensional materials has been long recognized as a powerful tool to create unique systems with properties that cannot be found in natural compounds \cite{Geim2013}. However, among the variety of vdW heterostructures and their various properties, only a few have revealed metallic and ferroelectric behaviour signatures\cite{Sharmaeaax5080}$^{,}$\cite{Fei2018}. Here we show ferroelectric semimetal made of double-gated double-layer graphene separated by an atomically thin crystal of hexagonal boron nitride, which demonstrating high room temperature mobility of the order of 10 m$^2$V$^{-1}$s$^{-1}$ and exhibits robust ambipolar switching in response to the external electric field. The observed hysteresis is tunable, reversible and persists above room temperature. Our fabrication method expands the family of ferroelectric vdW compounds and offers a route for developing novel phase-changing devices. 
\end{abstract}

\begin{document}

\flushbottom
\maketitle

\section*{Introduction}
Competing phases in condensed matter demonstrate the variety of physics phenomena:  superconductivity and ferromagnetism, charge density wave and superconductivity\cite{Xi2015} to name a few. The recently investigated ferroelectric semimetal such as WTe$_2$\cite{Sharmaeaax5080}$^{,}$\cite{Fei2018} as an example of an unintuitive interplay between polarization and free charge, which seems at first glance, should screen the former. Generally, metallic properties of a material do not favour macroscopic polarization, which means that ferroelectricity as a phenomenon is observed in many ferroelectrics of dielectric nature or semiconducting materials. Only a few examples exist for metallic ferroelectrics and even less number of candidates to show hysteretic behaviour\cite{Zhou_2020}. However, there is one experimental evidence of natural 2D ferroelectric semimetal at room temperature: a few-layer WTe$_2$\cite{Fei2018} made by exfoliation, and another example is artificially made 2D polar metal based on superlattices BaTiO$_3$/SrTiO$_3$/LaTiO$_3$ using advantage of molecular beam epitaxy growth\cite{osti_1466346}. Both of these materials demonstrate a room-temperature ferroelectric effect. A ferroelectric structural transition occurs in bulk (3D) LiOsO$_3$ crystals at 140K\cite{Shi2013}. The interest in these materials is hidden in the possibility of creating new quantum states, including coexisting ferroelectric, ferromagnetic, and superconducting phases\cite{Shi2013}, use them for functional nanoelectronics applications\cite{Sharmaeaax5080}. It is relevant to the combination of memory effects and conduction modulation, which improves transistor performance\cite{Liu2019}$^{,}$\cite{Si2019}.

Graphene has high-mobility charge carriers at room temperature with acoustic phonon scattering as a limiting factor\cite{Wang2013}. Bilayer graphene demonstrates  lower mobility at room temperature than monolayer graphene, but it is still much larger than any other semiconductors, and semimetals known\cite{Dean2010}$^{,}$\cite{Mayorov2011}$^{,}$\cite{Neto2009}. Therefore, it could improve the transport properties of ferroelectric metals based on vdW graphene heterostructures.  Previously a breakthrough idea of changing properties of two-dimensional materials by staking them in a different order is useful for constructing the new artificially made vdW heterostructures\cite{Geim2013}. For a ferroelectric metal to exist, several criteria need to be: second-order phase transition, breaking inversion symmetry, polarization switchability\cite{Zhou_2020}.   

Here we explore this general idea to create a recently discovered class of materials that combine semimetallic and ferroelectric properties in a single material\cite{PhysRevLett.115.087202}. The effect of the moiré pattern reveals the possibility of forming a strong ferroelectric characteristic for conventional 2D materials such as hexagonal boron nitride(hBN) and bilayer graphene\cite{Zheng2020}. Breaking of the inversion symmetry occurs by mechanical stacking of two individual layers under small angle rotation, which opens a new path to create different physical effects in graphene, besides the well know proximity of graphene to ferromagnets, superconductors, 2D materials with spin-orbit interactions. Here we study electronic properties and reproduce the ferroelectric effect in bilayer graphene intercalated with a monolayer hBN up to 325K and demonstrate a metal-to-insulator transition ferroelectric effect in graphene.

\begin{figure}[h]
		\centering
        \includegraphics[width=1.\textwidth]{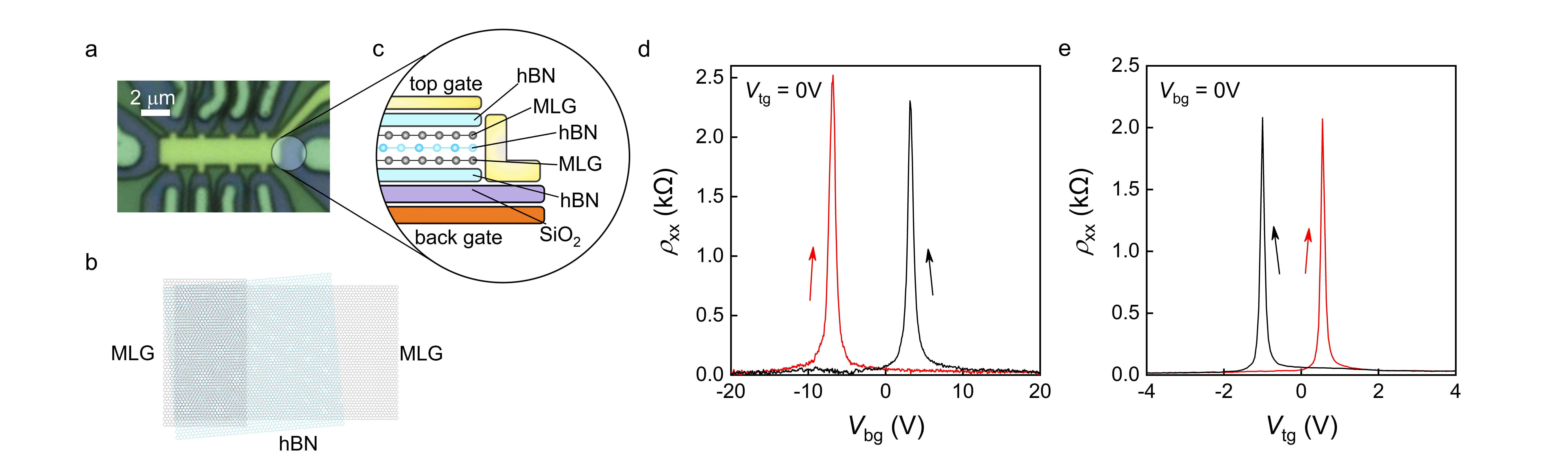}
	    \caption[width=0.8\textwidth]{Hexagonal boron nitride-separated quasi-twisted bilayer graphene. \textbf{a}. Optical photograph of our Hall bar device. An encapsulated qTBG heterostructure is connected to metal leads (dull green) and endowed with gold top gate(bright green) and bottom silicon gate electrodes. \textbf{b}. Schematic of the triple-layer structure. Two MLG layers are twisted by a small twist angle. \textbf{c}. The schematic of the qTBG device with top and bottom gates. \textbf{d} and \textbf{e}. The device's resistivity measured as a function of $V_{bg}$ and $V_{tg}$ at 2.1K for $V_{tg}=0$V and $V_{bg}=0$V, respectively.}
        \label{fig:Fig1}
    \hfill
\end{figure}

\section*{Fabrication}
Our device is a multiterminal Hall bar (Fig.\ref{fig:Fig1}\textbf{a}) made of quasi-twisted bilayer graphene (qTBG) using standard dry transfer technique\cite{Kretinin2014}. Graphene layers in such qTBG are separated by a monolayer of hexagonal boron nitride, allowing the graphene layers to be tunnelled transparent. The sandwich is encapsulated between two relatively thick hexagonal boron nitride slabs protecting graphene layers from the environment (Fig.\ref{fig:Fig1}\textbf{b}). The qTBG heterostructure is connected to metal contacts\cite{Wang2013} and endowed with top and bottom gate electrodes allowing independent control over the carrier density in each layer ($n_t$ and $n_b$, respectively) and relative displacement field between the layers. 

\section*{Results}
Electrical transport measurements are carried out in a $^3$He variable temperature inset system using the standard low-frequency lock-in technique. Fig.\ref{fig:Fig1}\textbf{c} shows our qTBG sample's resistivity as a function of back-gate voltage, $V_{bg}$, measured at T = 2.1K and zero top gate voltage, $V_{tg}$ = 0V. It exhibits familiar for high-mobility bilayer graphene behaviour with a sharp peak of about 2k$\Omega$ corresponding to the charge neutrality point (CNP), followed by a rapid decrease with increasing $V_{bg}$. When the gate voltage's sweep direction is reversed, the resistivity curve is shifted so that the CNP appears at 7 V, more than 10 V away from its initial position. The observed hysteresis is robust and reproduces itself for numerous gate (top and bottom) voltage sweep loops without an apparent sign of degradation of the hysteretic behaviour. A similar hysteresis is observed for the top gate sweep measured at $V_{bg}=0$V. Notice the CNP position, which is shifted to positive top gate voltages for the forward sweep and negative voltage for the backward sweep. 

\begin{figure}[h]
		\centering
        \includegraphics[width=0.8\textwidth]{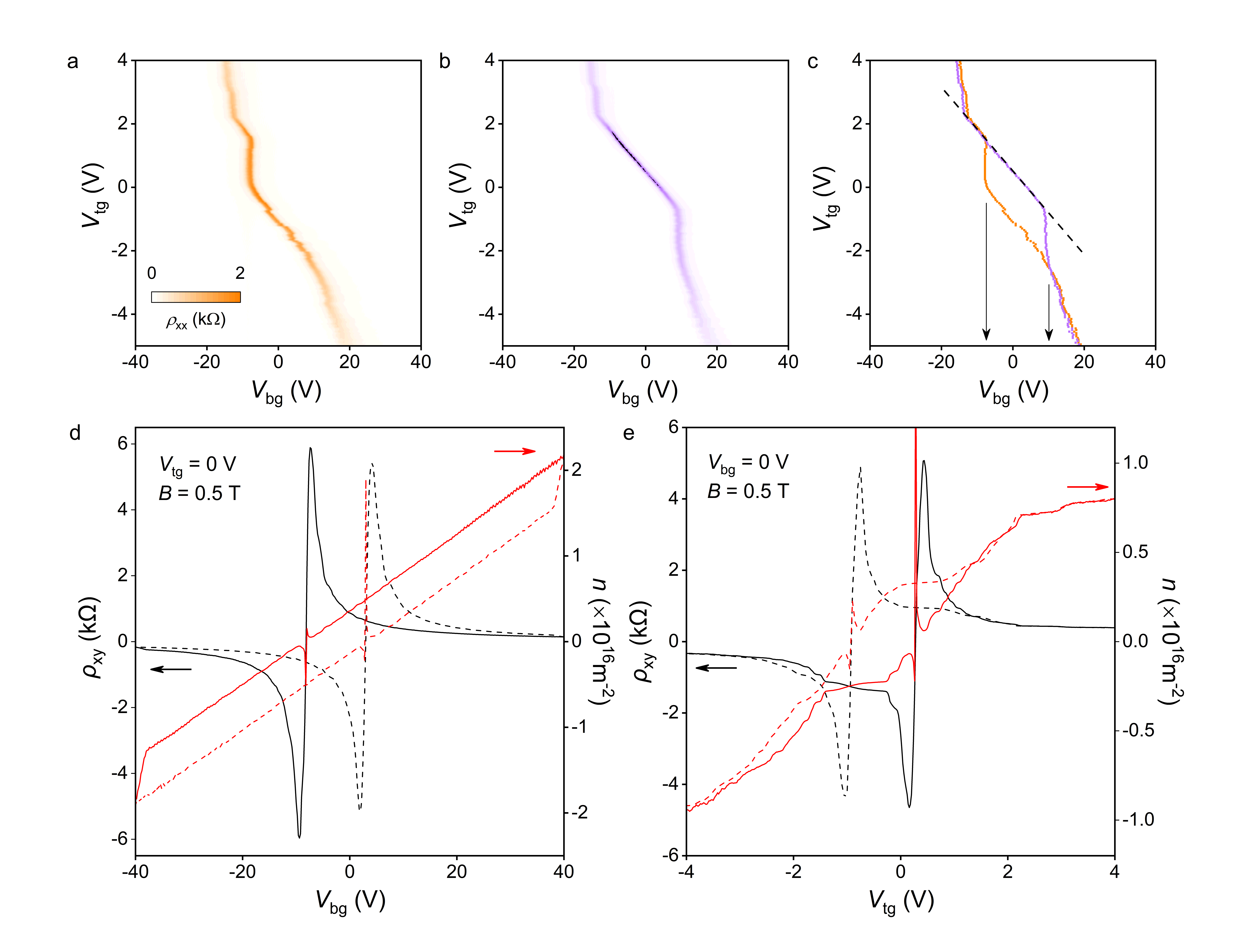}
	    \caption[width=0.8\textwidth]{Ferroelectric hysteresis at 2.1K. \textbf{a}-\textbf{b} The sample's resistivity as a function of the back gate and top gate voltages for the forward \textbf{a} and backward \textbf{b} back gate voltage sweeps at fixed $V_{tg}$. \textbf{c}. The maximum of the resistivity for the forward and backward as a function of the top gate and back gate voltages. The dashed line is the best fit for the forward and backward sweep's resistivity maxima in the linear ferroelectric regime. \textbf{d}. Hall effect measured at $B=0.5$T for forward (solid black curve) and backward (dashed black curve) sweeps as a function of back gate voltage at $V_{tg}=0$V. The red curves show corresponding concentrations calculated from the Hall voltage. \textbf{e}. Hall effect measured at $B=0.5$T for the forward (solid black curve) and backward (dashed black curve) sweeps as a function of top gate voltage at $V_{bg}=0$V. The red curves show corresponding concentrations calculated from the Hall voltage.}
        \label{fig:Fig2}
    \hfill
\end{figure}

We review observed hysteretic behaviour of resistivity of the qTBG intercalated with monolayer hBN, in more details. Previously, a moiré heterostructure hBN/bilayer graphene/hBN has demonstrated a similar hysteresis at low temperatures\cite{Zheng2020}. The resistivity map is measured as a function of the back gate voltage which charges from -40V to 40V for the forward sweep Fig.\ref{fig:Fig2}\textbf{a} and 40V to -40V for the backward sweep (Fig.\ref{fig:Fig2}\textbf{b}). Both maps are measured at 2.1K, and during the sweep, the top gate voltage is fixed. The top gate voltage changes from negative to positive values (from -5V to 4V). The dark region shows a noticeable difference in the resistivity peak position between the forward and backward sweeps. We plot Fig.\ref{fig:Fig2}\textbf{c} the maximum resistivity for the forward and backward back gate voltage sweeps to characterize the hysteresis. The hysteretic behaviour illustrates internal polarization in the heterostructure\cite{Zheng2020}, which changes the charge carrier concentration in both graphene layers. The external electric field created by the back and top gate voltages can reverse this polarization by applying a large back gate voltage ($|V_{bg}|>10$V) or top gate voltage $V_{tg}<-2.2$V and $V_{tg}>1.5$V. The peak shifts linearly in the range of $V_{bg}$ from -9.1V to 9.7V, which corresponds to a linear polarization for ferroelectrics\cite{lines1977principles}. The top gate's efficiency with respect to the bottom gate cannot be determined from the top and bottom hBN thicknesses (42.6nm and 111nm, respectively) using a small dielectric constant for hBN. If the relative dielectric constant for hBN is taken as 5, then the expected efficiency is about 11.6. However, the dashed lines in Fig.\ref{fig:Fig2}\textbf{c} correspond to the ratio of $V_{bg}/V_{tg}=7.6\pm0.1$. Therefore, the dielectric environment of graphene is significantly distorted by the ferroelectric effect. In the regions without hysteresis, the resistivity peak shift also does not correspond to the gate's efficiency. It enters the nonlinear polarization regime of ferroelectric, which is commonly attributed to the paraelectric phase\cite{lines1977principles}. In the paraelectric phase, the dielectric constant is not linear, and the resistivity of the structure depends strongly on the properties of hBN layers. To the best of our knowledge, this observation has not been reported before in 2D heterostructures.

The average concentration in the system then can be determined by the Hall effect (see supplementary). Calculations of the Hall concentration as a function of back-gate voltage at $V_{tg}=0$V are shown in Fig.\ref{fig:Fig2}\textbf{d}, \textbf{e}. The position of the maximum is shifted between the forward and backward shift on the same amount when the efficiency of the top gate is taken into account. We notice here that the view of this hysteresis and the positions of maxima and minima do not change if top gate voltage sweeps are used at fixed $V_{bg}$ (See supplementary).

Finally, we characterized the temperature dependence of the resistivity for the forward and backward sweeps. The peak resistivity is reducing with increasing temperature as expected for both monolayer and bilayer graphene, and the separation between peaks, which is related to spontaneous polarization in our structure, is decreasing with increasing temperature. Still, it does not disappear entirely, even at the maximum available temperature of 325K (Fig.\ref{fig:Fig3}\textbf{b}). The full data set is shown in the supplementary. The cross-over density between metallic ($d\rho/dT<0$) and insulator ($d\rho/dT>0$) states is equal to 5$\times$10$^{10}$ cm${^-}$${^2}$, which is in agreement with previous report\cite{Ponomarenko2011}.  The corresponding mobility is found to be better than 10 m$^{2}$V$^{-1}$s$^{-1}$ at room temperature (Fig.\ref{fig:Fig3}\textbf{d}), which is higher than in bilayer graphene\cite{Schmitz2017} and comparable with acoustic phonon limited mobility in monolayer graphene\cite{Wang2013}. The charge carrier mobility reduces with increasing temperature, as shown in Fig.\ref{fig:Fig3}\textbf{d} linearly.

\begin{figure}[h]
	\centering
	\includegraphics[width=0.6\textwidth]{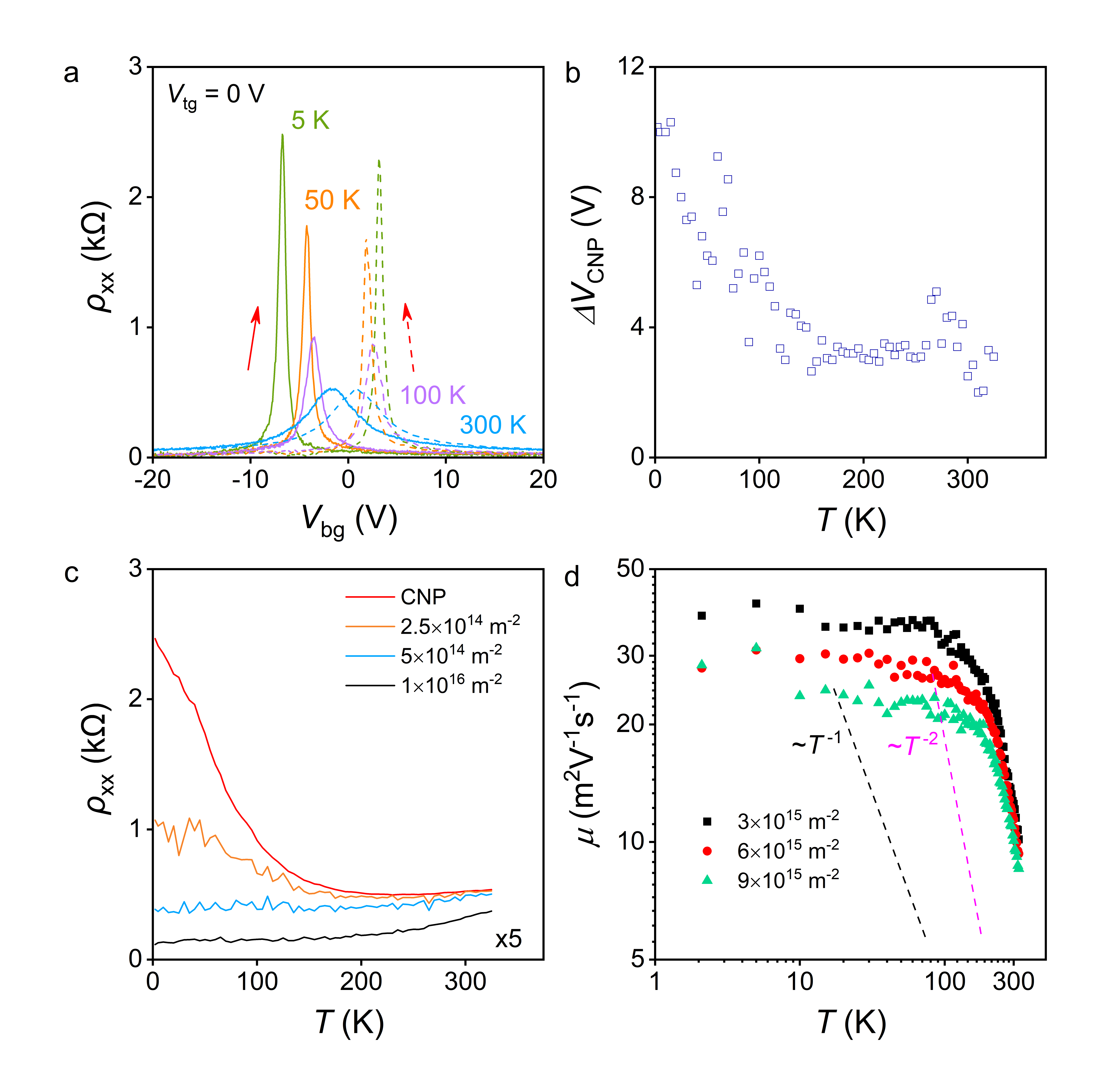}
	\caption[width=0.8\textwidth]{Transport properties at high temperatures. \textbf{a}. Temperature dependence of the resistivity as a function of back gate voltage at $V_{tg}=0$V for the forward (solid style) and backward sweeps (dashed style) for four selected temperatures. \textbf{b}. The voltage difference between CNP positions of the backward and forward sweeps. The temperature changes from 2K to 325K. \textbf{c}. The temperature dependence of the resistivity for the forward sweeps for different electron concentrations. The smallest resistivity is multiplied by 5. \textbf{d}. The mobility of electron gas as a function of temperature measured at 3$\times10^{15}$ m$^{-2}$, 6$\times10^{15}$ m$^{-2}$ and 9$\times10^{15}$ m$^{-2}$. The dashed lines are guides for eyes.}
	\label{fig:Fig3}
	\hfill
\end{figure}

\section*{Discussion}
Previously a strong ferroelectric effect was observed in aligned and rotated by 30$^{\circ}$ hBN/bilayer graphene heterostructures\cite{Zheng2020}. The authors argued that the different parts of a supercell could induce spontaneous polarization in such a structure. Here we study a similar system without intentional alignment between hBN and graphene layers. The absence of any supercell at low energy could be justified by the gate voltage dependence of the resistance. In this case, the band structure should have more features at low energy, which creates extra CNPs as was reported in \cite{Wang2019}.  Also, Brown-Zak oscillations\cite{Kumar2017}, which could be an indication of alignment between graphene and hBN, are not observed in our device. Therefore, our system's ferroelectric effect could not be of the same origin as in the intentionally aligned graphene/hBN structure\cite{Zheng2020}. However, we could not exclude 30$^{\circ}$ rotation\cite{Zheng2020} or the inversion symmetry breaking in the vertical graphene/hBN/graphene heterostructure as a cause of ferroelectricity. 
The twisted hBN layers can demonstrate the ferroelectric effect\cite{Woods2021}. This observation was reported in \cite{Stern2020},\cite{Yasuda2020}. If we assume that the monolayer hBN can be easily twisted when placed between two graphene layers. In that case, the polarization can be created by the top hBN and monolayer hBN or between monolayer hBN and bottom hBN layers. This effect needs further investigation.

Mobility of the charge carriers in graphene at high temperature is limited by acoustic phonon scattering\cite{Wang2013}. The temperature dependence of mobility is inversely proportional to temperature. This observation is valid for graphene and its bilayer\cite{Min2011}. The temperature dependence of resistivity is proportional to $\rho\propto T$ in the case of twisted bilayer graphene, which was observed experimentally\cite{Polshyn2019} and predicted theoretically\cite{Wu2019} for small angles of rotation. Our observations demonstrate inverse parabolic dependence of the mobility $\rho\propto T^{-2}$ as shown in Fig.\ref{fig:Fig3}\textbf{d}, which we attribute to the contribution of optical phonons in hBN substrate\cite{Tan2013} or polar optical phonons\cite{Li2010} in agreement with ferroelectric nature of our structure. The low-temperature mobility is limited by scattering at the edges of the sample.\\

\section*{Methods}
\subsection{Transport measurements}
The device was measured in an Oxford Instruments TeslatronPT cryogen-free superconducting magnet system equipped with Oxford Instruments HelioxVT Sorption Pumped $^3$He Refrigerator insert (300mK,14T) and the magnetic field applied perpendicular to the plane of the film. Stanford Research Systems SR830 lock-in is used to apply a AC bias current with a 100MOhm bias resistor at a frequency of 13.333 Hz, and Keithley 2614B SourceMeters were used to apply DC current with a 100MOhm bias resistor. Keithley 2400 SourceMeters were used to apply voltages to the gates.

\subsection{AFM and Raman measurements}
AFM measurements are performed with a Bruker Dimension Fastscan system at tapping mode. Scan area of the bottom/top hBN are shown in the right/left red dotted boxes in the supplementary. The length to width ratio is 1.16. The width of the sample is 2.7$\mu$m. The top and bottom hBN thicknesses are equal to 42.6nm and 111nm, respectively.

Room temperature Raman scattering is performed using a WITec/alpha 300R confocal microscope with a 532nm laser under ambient conditions. The laser power was kept below 1mW to avoid damage or heating. The G and 2D peaks in the Raman spectra are fitted with Lorentzian. Typical Raman spectra of different positions of the heterostructure are plotted in the Supplementary.

\bibliography{sample}

\begin{thebibliography}{10}
\urlstyle{rm}
\expandafter\ifx\csname url\endcsname\relax
  \def\url#1{\texttt{#1}}\fi
\expandafter\ifx\csname urlprefix\endcsname\relax\def\urlprefix{URL }\fi
\expandafter\ifx\csname doiprefix\endcsname\relax\def\doiprefix{DOI: }\fi
\providecommand{\bibinfo}[2]{#2}
\providecommand{\eprint}[2][]{\url{#2}}

\bibitem{Geim2013}
\bibinfo{author}{Geim, A.~K.} \& \bibinfo{author}{Grigorieva, I.~V.}
\newblock \bibinfo{journal}{\bibinfo{title}{Van der waals heterostructures}}.
\newblock {\emph{\JournalTitle{Nature}}} \textbf{\bibinfo{volume}{499}},
  \bibinfo{pages}{419–425},
  \doiprefix\url{https://www.nature.com/articles/nature12385}
  (\bibinfo{year}{2013}).

\bibitem{Sharmaeaax5080}
\bibinfo{author}{Sharma, P.} \emph{et~al.}
\newblock \bibinfo{journal}{\bibinfo{title}{A room-temperature ferroelectric
  semimetal}}.
\newblock {\emph{\JournalTitle{Science Advances}}}
  \textbf{\bibinfo{volume}{5}}, \bibinfo{pages}{eaax5080},
  \doiprefix\url{https://advances.sciencemag.org/content/5/7/eaax5080}
  (\bibinfo{year}{2019}).

\bibitem{Fei2018}
\bibinfo{author}{Fei, Z.} \emph{et~al.}
\newblock \bibinfo{journal}{\bibinfo{title}{Ferroelectric switching of a
  two-dimensional metal}}.
\newblock {\emph{\JournalTitle{Nature}}} \textbf{\bibinfo{volume}{560}},
  \bibinfo{pages}{336–339},
  \doiprefix\url{https://doi.org/10.1038/s41586-018-0336-3}
  (\bibinfo{year}{2018}).

\bibitem{Xi2015}
\bibinfo{author}{Xi, X.} \emph{et~al.}
\newblock \bibinfo{journal}{\bibinfo{title}{Strongly enhanced
  charge-density-wave order in monolayer nbse$_2$}}.
\newblock {\emph{\JournalTitle{Nature Nanotech.}}}
  \textbf{\bibinfo{volume}{10}}, \bibinfo{pages}{765–769},
  \doiprefix\url{https://doi.org/10.1038/nnano.2015.143}
  (\bibinfo{year}{2015}).

\bibitem{Zhou_2020}
\bibinfo{author}{Zhou, W.~X.} \& \bibinfo{author}{Ariando, A.}
\newblock \bibinfo{journal}{\bibinfo{title}{Review on ferroelectric/polar
  metals}}.
\newblock {\emph{\JournalTitle{Japanese Journal of Applied Physics}}}
  \textbf{\bibinfo{volume}{59}}, \bibinfo{pages}{SI0802},
  \doiprefix\url{https://doi.org/10.35848/1347-4065/ab8bbf}
  (\bibinfo{year}{2020}).

\bibitem{osti_1466346}
\bibinfo{author}{Cao, Y.} \emph{et~al.}
\newblock \bibinfo{journal}{\bibinfo{title}{Artificial two-dimensional polar
  metal at room temperature}}.
\newblock {\emph{\JournalTitle{Nature Communications}}}
  \textbf{\bibinfo{volume}{9}}, \bibinfo{pages}{1547},
  \doiprefix\url{10.1038/s41467-018-03964-9} (\bibinfo{year}{2018}).

\bibitem{Shi2013}
\bibinfo{author}{Shi, Y.} \emph{et~al.}
\newblock \bibinfo{journal}{\bibinfo{title}{A ferroelectric-like structural
  transition in a metal}}.
\newblock {\emph{\JournalTitle{Nature Materials}}}
  \textbf{\bibinfo{volume}{12}}, \bibinfo{pages}{1024--1027},
  \doiprefix\url{https://doi.org/10.1038/nmat3754} (\bibinfo{year}{2013}).

\bibitem{Liu2019}
\bibinfo{author}{Liu, X.} \emph{et~al.}
\newblock \bibinfo{journal}{\bibinfo{title}{Vertical ferroelectric switching by
  in-plane sliding of two-dimensional bilayer wte$_2$}}.
\newblock {\emph{\JournalTitle{Nanoscale}}} \textbf{\bibinfo{volume}{11}},
  \bibinfo{pages}{18575--18581},
  \doiprefix\url{https://doi.org/10.1039/C9NR05404A} (\bibinfo{year}{2019}).

\bibitem{Si2019}
\bibinfo{author}{Si, M.} \emph{et~al.}
\newblock \bibinfo{journal}{\bibinfo{title}{A ferroelectric semiconductor
  field-effect transistor}}.
\newblock {\emph{\JournalTitle{Nature Nanoelectronics}}}
  \textbf{\bibinfo{volume}{2}}, \bibinfo{pages}{580–586},
  \doiprefix\url{https://doi.org/10.1038/s41928-019-0338-7}
  (\bibinfo{year}{2019}).

\bibitem{Wang2013}
\bibinfo{author}{Wang, L.} \emph{et~al.}
\newblock \bibinfo{journal}{\bibinfo{title}{One-dimensional electrical contact
  to a two-dimensional material}}.
\newblock {\emph{\JournalTitle{Science}}} \textbf{\bibinfo{volume}{342}},
  \bibinfo{pages}{614--617},
  \doiprefix\url{https://doi.org/10.1126/science.1244358}
  (\bibinfo{year}{2013}).

\bibitem{Dean2010}
\bibinfo{author}{Dean, C.~R.} \emph{et~al.}
\newblock \bibinfo{journal}{\bibinfo{title}{Boron nitride substrates for
  high-quality graphene electronics}}.
\newblock {\emph{\JournalTitle{Nature Nanotechnology}}}
  \textbf{\bibinfo{volume}{5}}, \bibinfo{pages}{722–726},
  \doiprefix\url{https://doi.org/10.1038/nnano.2010.172}
  (\bibinfo{year}{2010}).

\bibitem{Mayorov2011}
\bibinfo{author}{Mayorov, A.~S.} \emph{et~al.}
\newblock \bibinfo{journal}{\bibinfo{title}{Micrometer-scale ballistic
  transport in encapsulated graphene at room temperature}}.
\newblock {\emph{\JournalTitle{Nano Lett.}}} \textbf{\bibinfo{volume}{11}},
  \bibinfo{pages}{2396–2399},
  \doiprefix\url{https://doi.org/10.1021/nl200758b} (\bibinfo{year}{2011}).

\bibitem{Neto2009}
\bibinfo{author}{Castro~Neto, A.~H.}, \bibinfo{author}{Guinea, F.},
  \bibinfo{author}{Peres, N. M.~R.}, \bibinfo{author}{Novoselov, K.~S.} \&
  \bibinfo{author}{Geim, A.~K.}
\newblock \bibinfo{journal}{\bibinfo{title}{The electronic properties of
  graphene}}.
\newblock {\emph{\JournalTitle{Rev. Mod. Phys.}}}
  \textbf{\bibinfo{volume}{81}}, \bibinfo{pages}{109--162},
  \doiprefix\url{https://link.aps.org/doi/10.1103/RevModPhys.81.109}
  (\bibinfo{year}{2009}).

\bibitem{PhysRevLett.115.087202}
\bibinfo{author}{Puggioni, D.}, \bibinfo{author}{Giovannetti, G.},
  \bibinfo{author}{Capone, M.} \& \bibinfo{author}{Rondinelli, J.~M.}
\newblock \bibinfo{journal}{\bibinfo{title}{Design of a mott multiferroic from
  a nonmagnetic polar metal}}.
\newblock {\emph{\JournalTitle{Phys. Rev. Lett.}}}
  \textbf{\bibinfo{volume}{115}}, \bibinfo{pages}{087202},
  \doiprefix\url{https://link.aps.org/doi/10.1103/PhysRevLett.115.087202}
  (\bibinfo{year}{2015}).

\bibitem{Zheng2020}
\bibinfo{author}{Zheng, Z.} \emph{et~al.}
\newblock \bibinfo{journal}{\bibinfo{title}{Unconventional ferroelectricity in
  moiré heterostructures}}.
\newblock {\emph{\JournalTitle{Nature}}} \textbf{\bibinfo{volume}{588}},
  \bibinfo{pages}{71–76},
  \doiprefix\url{https://doi.org/10.1038/s41586-020-2970-9}
  (\bibinfo{year}{2020}).

\bibitem{Kretinin2014}
\bibinfo{author}{Kretinin, A.~V.} \emph{et~al.}
\newblock \bibinfo{journal}{\bibinfo{title}{Electronic properties of graphene
  encapsulated with different two-dimensional atomic crystals}}.
\newblock {\emph{\JournalTitle{Nano Lett.}}} \textbf{\bibinfo{volume}{14}},
  \bibinfo{pages}{3270–3276},
  \doiprefix\url{https://doi.org/10.1021/nl5006542} (\bibinfo{year}{2014}).

\bibitem{lines1977principles}
\bibinfo{author}{Lines, M.}
\newblock \emph{\bibinfo{title}{Principles and applications of ferroelectrics
  and related materials}} (\bibinfo{publisher}{Clarendon Press},
  \bibinfo{address}{Oxford England}, \bibinfo{year}{1977}).

\bibitem{Ponomarenko2011}
\bibinfo{author}{Ponomarenko, L.~A.} \emph{et~al.}
\newblock \bibinfo{journal}{\bibinfo{title}{Tunable metal–insulator
  transition in double-layer graphene heterostructures}}.
\newblock {\emph{\JournalTitle{Nature Phys.}}} \textbf{\bibinfo{volume}{7}},
  \bibinfo{pages}{958–961}, \doiprefix\url{https://doi.org/10.1038/nphys2114}
  (\bibinfo{year}{2011}).

\bibitem{Schmitz2017}
\bibinfo{author}{Schmitz, M.} \emph{et~al.}
\newblock \bibinfo{journal}{\bibinfo{title}{High mobility dry-transferred cvd
  bilayer graphene}}.
\newblock {\emph{\JournalTitle{Appl. Phys. Lett.}}}
  \textbf{\bibinfo{volume}{110}}, \bibinfo{pages}{263110},
  \doiprefix\url{https://doi.org/10.1063/1.4990390} (\bibinfo{year}{2017}).

\bibitem{Wang2019}
\bibinfo{author}{Wang, Z.} \emph{et~al.}
\newblock \bibinfo{journal}{\bibinfo{title}{Composite super-moiré lattices in
  double-aligned graphene heterostructures}}.
\newblock {\emph{\JournalTitle{Science Advances}}}
  \textbf{\bibinfo{volume}{5}}, \bibinfo{pages}{eaay8897},
  \doiprefix\url{https://doi.org/10.1126/sciadv.aay8897}
  (\bibinfo{year}{2019}).

\bibitem{Kumar2017}
\bibinfo{author}{Kumar, R.~K.} \emph{et~al.}
\newblock \bibinfo{journal}{\bibinfo{title}{High-temperature quantum
  oscillations caused by recurring bloch states in graphene superlatticess}}.
\newblock {\emph{\JournalTitle{Science}}} \textbf{\bibinfo{volume}{357}},
  \bibinfo{pages}{181--184},
  \doiprefix\url{https://doi.org/10.1126/science.aal3357}
  (\bibinfo{year}{2017}).

\bibitem{Woods2021}
\bibinfo{author}{Woods, C.~R.} \emph{et~al.}
\newblock \bibinfo{journal}{\bibinfo{title}{Charge-polarized interfacial
  superlattices in marginally twisted hexagonal boron nitride}}.
\newblock {\emph{\JournalTitle{Nature Communications}}}
  \textbf{\bibinfo{volume}{12}}, \bibinfo{pages}{347},
  \doiprefix\url{https://doi.org/10.1038/s41467-020-20667-2}
  (\bibinfo{year}{2021}).

\bibitem{Stern2020}
\bibinfo{author}{Stern, M.~V.} \emph{et~al.}
\newblock \bibinfo{journal}{\bibinfo{title}{Interfacial ferroelectricity by van
  der waals sliding}}.
\newblock {\emph{\JournalTitle{arXiv:2010.05182}}}  (\bibinfo{year}{2020}).

\bibitem{Yasuda2020}
\bibinfo{author}{Yasuda, K.}, \bibinfo{author}{Wang, X.},
  \bibinfo{author}{Watanabe, K.}, \bibinfo{author}{Taniguchi, T.} \&
  \bibinfo{author}{Jarillo-Herrero, P.}
\newblock \bibinfo{journal}{\bibinfo{title}{Stacking-engineered
  ferroelectricity in bilayer boron nitride}}.
\newblock {\emph{\JournalTitle{arXiv:2010.06600}}}  (\bibinfo{year}{2020}).

\bibitem{Min2011}
\bibinfo{author}{Min, H.}, \bibinfo{author}{Hwang, E.~H.} \&
  \bibinfo{author}{Das~Sarma, S.}
\newblock \bibinfo{journal}{\bibinfo{title}{Chirality-dependent phonon-limited
  resistivity in multiple layers of graphene}}.
\newblock {\emph{\JournalTitle{Phys. Rev. B}}} \textbf{\bibinfo{volume}{83}},
  \bibinfo{pages}{161404},
  \doiprefix\url{https://link.aps.org/doi/10.1103/PhysRevB.83.161404}
  (\bibinfo{year}{2011}).

\bibitem{Polshyn2019}
\bibinfo{author}{Polshyn, H.} \emph{et~al.}
\newblock \bibinfo{journal}{\bibinfo{title}{Large linear-in-temperature
  resistivity in twisted bilayer graphene}}.
\newblock {\emph{\JournalTitle{Nature Phys.}}} \textbf{\bibinfo{volume}{15}},
  \bibinfo{pages}{1011–1016},
  \doiprefix\url{https://doi.org/10.1038/s41567-019-0596-3}
  (\bibinfo{year}{2019}).

\bibitem{Wu2019}
\bibinfo{author}{Wu, F.}, \bibinfo{author}{Hwang, E.} \&
  \bibinfo{author}{Das~Sarma, S.}
\newblock \bibinfo{journal}{\bibinfo{title}{Phonon-induced giant linear-in-$t$
  resistivity in magic angle twisted bilayer graphene: Ordinary strangeness and
  exotic superconductivity}}.
\newblock {\emph{\JournalTitle{Phys. Rev. B}}} \textbf{\bibinfo{volume}{99}},
  \bibinfo{pages}{165112},
  \doiprefix\url{https://link.aps.org/doi/10.1103/PhysRevB.99.165112}
  (\bibinfo{year}{2019}).

\bibitem{Tan2013}
\bibinfo{author}{Lin, I.-T.} \& \bibinfo{author}{Liu, J.-M.}
\newblock \bibinfo{journal}{\bibinfo{title}{Surface polar optical phonon
  scattering of carriers in graphene on various substrates}}.
\newblock {\emph{\JournalTitle{Appl. Phys. Lett.}}}
  \textbf{\bibinfo{volume}{103}}, \bibinfo{pages}{081606},
  \doiprefix\url{https://doi.org/10.1063/1.4819395} (\bibinfo{year}{2013}).

\bibitem{Li2010}
\bibinfo{author}{Li, X.}, \bibinfo{author}{Barry, E.~A.},
  \bibinfo{author}{Zavada, v.}, \bibinfo{author}{Nardelli, M.~B.} \&
  \bibinfo{author}{Kim, K.~W.}
\newblock \bibinfo{journal}{\bibinfo{title}{Surface polar phonon dominated
  electron transport in graphene}}.
\newblock {\emph{\JournalTitle{Appl. Phys. Lett.}}}
  \textbf{\bibinfo{volume}{97}}, \bibinfo{pages}{232105},
  \doiprefix\url{https://doi.org/10.1063/1.3525606} (\bibinfo{year}{2010}).

\end{thebibliography}

\section*{Acknowledgements}

We thank B.G. Wang, K.S. Novoselov for useful discussions. This work is supported by the National Key R$\&$D Program of China (grant nos. SQ2018YFA030066, SQ2018YFA030143), the National Natural Science Foundation of China (no. 11974169) and the Fundamental Research Funds for the Central Universities (nos. 020414380087, 020414913201), and the Basic Research Program of Jiangsu Province (Grant No. BK20190283).

\section*{Author contributions statement}

A.S.M. and G.Y. designed the project. Y.W. and fabricated the samples, S.J. and J.X. performed transport measurements, X.C. and G.M. did the AFM and Raman research, K.W. and T.T. provided hBN crystals. S. J., J.X., Y.W., and A.S.M performed data analysis, D.Z., P.W., G.M., Y.H, J.H., and A.S.M provided the experimental support. Y.W., A.S.M and G.Y. wrote the manuscript. All authors participated in the discussions. 

\section*{Competing interests}
The authors declare no competing interests.

\section*{Additional information}
\textbf{Supplementary information} is available for this paper at <>.

\textbf{Correspondence and requests for materials} should be addressed to Y.W, A.S.M. or G.Y.

\end{document}